\documentclass[prb,twocolumn,floatfix]{revtex4}

\usepackage{graphics}
\usepackage{graphicx}
\usepackage{amssymb}
\usepackage{amsmath}
\usepackage{latexsym}
\usepackage{bm}
\usepackage[T1]{fontenc}

\begin{document}
\title{Ising Like Order by Disorder In The Pyrochlore Antiferromagnet with Dzyaloshinskii-Moriya Interactions}
\author{B. Canals, M. Elhajal, C. Lacroix}
\affiliation{$^1$Institut NEEL, CNRS \& Université Joseph Fourier,
BP 166, F-38042 Grenoble Cedex 9\\}

\begin{abstract}
It is shown that the mechanism of order out of disorder is at work
in the antisymmetric pyrochlore antiferromagnet.
Quantum as well as thermal fluctuations break the continuous
degeneracy of the classical ground state manifold and reduce its
symmetry to $\mathbb{Z}_3 \times \mathbb{Z}_2$.
The role of anisotropic symmetric exchange is also investigated
and we conclude that this discrete like ordering is robust with
respect to these second order like interactions.
The antisymmetric pyrochlore antiferromagnet is therefore expected
to order at low temperatures, whatever the symmetry type of its
interactions, in both the classical and semi classical limits.
\end{abstract}

\pacs{75.10.Hk, 75.30.Et, 75.30.Gw, 75.50.Ee}

\maketitle

\section{INTRODUCTION}

The role of geometrical frustration in magnetic systems is at
present one of the open question in the physics of strongly
correlated systems since it can lead to novel low temperature
behaviors.
One of the most studied model is the nearest neighbor
antiferromagnetic Heisenberg model on the pyrochlore lattice.
For
classical spins, the ground state is known to have macroscopic
degeneracy, preventing any magnetic ordering at T=0.
In real
systems, this degeneracy is often removed, partially or totally,
by additional interactions such as dipolar
interactions\cite{Palmer}, 2nd neighbor
exchange\cite{Reimers1992}, single ion
anisotropy\cite{Harris1997}, magneto-elastic
coupling\cite{Tchernyshov}...
This explains why pyrochlore compounds often order at low
temperature, but their ordering temperature is usually  much
smaller than the paramagnetic Curie temperature $\Theta_{p}$ since
$T_{N}$ is not related to the exchange interaction J, but to the
additional small interaction.
Another type of process which can remove the degeneracy is order
by disorder: as proposed first by Villain\cite{Villain1979} the
classical degeneracy can be lifted by fluctuations, if
fluctuations around these classical ground states select one of
these states.
Both classical and quantum fluctuations are able to
produce order by disorder, trough their contribution to entropy or
zero point energy respectively.
In this paper we study the effect
of Dzyaloshinsky-Moriya interactions in the Heisenberg pyrochlore
model.
In an earlier paper \cite{elhajal2005}, we
have shown that the T=0 K degeneracy is completly removed only for
one particular choice of the DM vectors.
Here we show that if the
DM vectors are such that the T=0 degeneracy is not removed, both
classical and quantum fluctuations select 6 equivalent ordered
configurations,leading to a well defined magnetic structure.
\begin{figure}
\includegraphics[width=75mm]{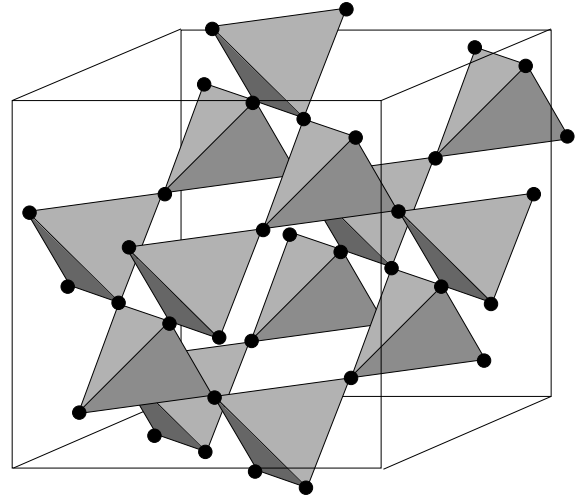}
\caption{\label{fig:fig1}The pyrochlore lattice. The interplay of
the frustration of the tetrahedral unit cell with the weak
connectivity of the (corner sharing) tetrahedra provides peculiar
magnetic properties to the pyrochlore antiferromagnet.}
\end{figure}

\section{DMI AND MORIYA'S RULES FOR THE PYROCHLORE LATTICE}
\label{sec:DMI-rules}

Dzyaloshinsky has shown\cite{Dzyaloshinsky} that, in crystals with
no inversion center, the usual isotropic exchange
$J\,\vec{S_{i}}.\vec{S_{j}}$ is not the only magnetic interaction
and antisymmetric exchange
$\vec{D}_{ij}.(\vec{S_{i}}\times\vec{S_{j}})$ is allowed.
In magnetic oxydes, exchange interactions are usually attributed
to superexchange mechanism, which involves virtual hopping between
2 neighboring magnetic ions.
Taking into account Coulomb
repulsion and Pauli's principle,Anderson  has explicited a
microscopic mecanism which leads to isotropic super-exchange
interactions.
Later, Moriya has shown\cite{MoriyaPR,MoriyaPRL}
that inclusion of spin-orbit coupling  on the magnetic ions in 1st
and 2nd order leads to antisymmetric exchange and anisotropic
exchange respectively.
Moriya's microscopic derivation of the DMI is only valid for
insulators but other possible microscopic mecanism relevant for
other materials were explicited, for instance in systems with RKKY
interactions\cite{Fert,Levy}.

Whatever the microscopic origin of the DMI is, there are always symmetry
constraints on the possible \textbf{D} vectors which may appear in
the hamiltonian.
Indeed, the hamiltonian must be invariant under the symmetry
operations of the crystal and this will restrict the possible
\textbf{D} vectors to those for which
$\vec{D}_{ij}.(\vec{Si}\times\vec{Sj})$ is invariant under these
symmetry operations.
This way of constraining the \textbf{D}
vectors has been given the name of Moriya's rules.
Using these
rules, we have shown\cite{elhajal2005} that two types of DM's interactions are
compatible with the pyrochlore space group symmetry.
These two
patterns of \textbf{D} vectors have been given the name of
"direct" and "indirect" case and are described in figure
\ref{fig:fig2}.

As shown in our previous work, the classical
ground state is always $q=0$.
It is non degenerate in the direct
case and is described by the so called all-in all-out magnetic
phase; this case will not be discussed in this paper.
From now on, we will focus on the second case of indirect DMI,
where the ground state manifold is reduced compared to the case
with no DMI, but is still continuously degenerate as shown in
Ref.~\onlinecite{elhajal2005}.
The role of
anisotropic exchange will also be discussed in this paper.

\section{ROLE OF FLUCTUATIONS WITHIN THE GROUND STATE MANIFOLD}
\label{sec:fluctuations}

From now on, we will focus on the second case of indirect DMI,
where the ground state manifold is $q=0$, but is continuously
degenerate.
There are two kinds of ground states\cite{elhajal2005} :
(i) phases can be coplanar, belonging either
to the $xy$, $yz$ or $zx$ plane and consists in perpendicular
pairs of anticollinear spins (see figure \ref{fig:fig7})which
can rotate freely within these planes.
\begin{figure}
\includegraphics[width=85mm]{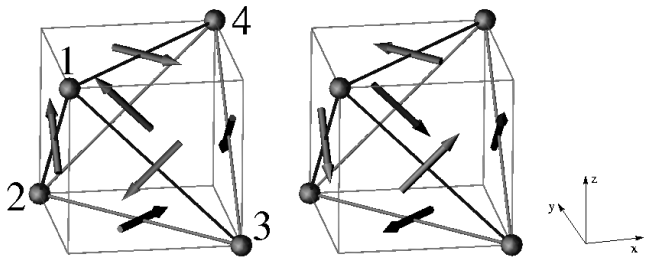}
\caption{\label{fig:fig2}\textbf{D} vectors for the DMI in the
pyrochlore lattice. The convention is taken to fix the order for
the cross products (always
$\mathbf{D}.\mathbf{S}_i\times\mathbf{S}_j$ with $j>i$). The two
possible DMI are those obtained by varying the direction of the
\textbf{D} vectors ($\mathbf{D}\to-\mathbf{D}$). The case with the
\textbf{D} as represented on the left is refered to as the
``direct'' case and the other case (on the right) is the
``indirect'' case. Once the DMI between two spins is fully
specified, the others DMI are also fixed and obtained by applying
the different $\frac{2\pi}{3}$ rotations around the cube's
diagonals which leave the tetrahedron invariant. The DMI in the
rest of the lattice are also fixed and obtained by applying
appropriate symmetry operations of the pyrochlore lattice.}
\end{figure}
\begin{figure}
\includegraphics[width=80mm]{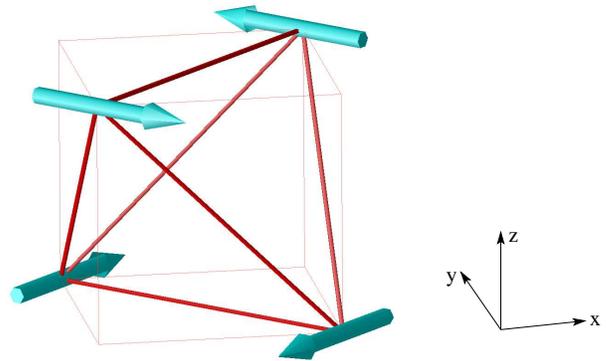}
\vskip15mm \caption{\label{fig:fig7}One possible ground state in
the case of indirect DMI. The ground state for the whole
pyrochlore lattice is a $\mathbf{q}=\mathbf{0}$ structure so that
only one tetrahedron is represented. Similar structures in the
$zx$ and $yz$ planes are degenerate.}
\end{figure}
(ii) There are also non coplanar phases, parametrized by one angle $\varphi$
as follow,
\begin{displaymath}
\begin{array}{c}
\label{eq:ground-state}
\mathbf{S}_{1}=\left\{
\begin{array}{l}
\cos\theta\cos\left(\varphi-\frac{\pi}{4}\right)\\
\cos\theta\sin\left(\varphi-\frac{\pi}{4}\right)\\
\sin\left(\theta\right)
\end{array}
\right. \mathbf{S}_{2}=\left\{
\begin{array}{l}
\cos\theta\cos\left(-\varphi+\frac{\pi}{4}\right)\\
\cos\theta\sin\left(-\varphi+\frac{\pi}{4}\right)\\
-\sin\left(\theta\right)
\end{array}
\right.\\
~~\\
\mathbf{S}_{3}=\left\{
\begin{array}{l}
\cos\theta\cos\left(-\varphi-\frac{3\pi}{4}\right)\\
\cos\theta\sin\left(-\varphi-\frac{3\pi}{4}\right)\\
-\sin\left(\theta\right)
\end{array}
\right. \mathbf{S}_{4}=\left\{
\begin{array}{l}
\cos\theta\cos\left(\varphi+\frac{3\pi}{4}\right)\\
\cos\theta\sin\left(\varphi+\frac{3\pi}{4}\right)\\
\sin\left(\theta\right)
\end{array}
\right.
\end{array}
\end{displaymath}

\noindent where the spins are labeled as in figure \ref{fig:fig2}
and where $\varphi$ and $\theta$ are \emph{not} independant~:
\begin{equation}
\theta=\arctan(\sqrt{2}\sin\varphi) \label{eqn:11h36}
\end{equation}
Consequently, it is possible to picture the classical ground state
manifold by following the paths defined by the arrow of say, spin
1, in all allowed configurations (actually, fixing any spin
determines the configuration of the 3 other spins).
Such paths are depicted in figure \ref{fig:spin1-path}.
\begin{figure}
\vskip0.5cm
\includegraphics[width=80mm]{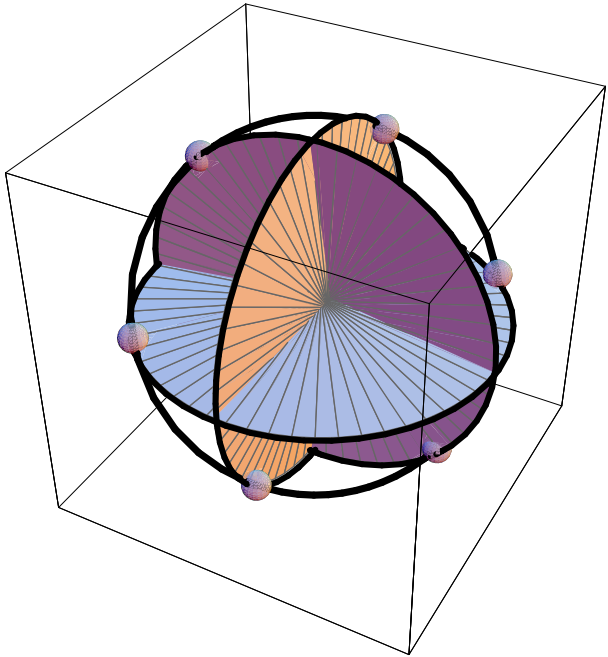}
\caption{\label{fig:spin1-path}Description of the one dimensional
ground state manifold.
Any move within the ground state can be described by the possible
path followed by only one of the four spin as they are all
parametrized by the same angle (see Eq.~\ref{eqn:11h36}).
Here, only spin 1 is described.
It can belong either to the $xy$, $yz$ or $zx$ plane, or be in a
non coplanar phase.
These non coplanar phases intersect coplanar phases at 6 points
and the {\it only} way to go from one plane to another is to pass
through these intersections, depicted by spheres.
}
\end{figure}
The ground state is therefore a multiply arcwise connected one
dimensional manifold, represented by the thick black lines on
figure \ref{fig:spin1-path}.
Six configurations play a particular role, since they are at the
intersection of two lines, i.e they belong to both kinds of
ground states defined above, where spins are
coplanar, either in the $xy$, $yz$ or $zx$ plane (see
Fig.~\ref{fig:fig7}).
In these phases, take for instance the $xy$ plane, two spins are
anti parallel (aligned along the (110) axis) and perpendicular to
the two remaining spins (aligned along the (1$\bar{1}$0) axis).
These configurations, related by time reversal symmetry and
rotations, are globally invariant under $\mathbb{Z}_3 \times
\mathbb{Z}_2$, and will be referred to this way from now on.
It is of practical and theoretical interest to test wether
fluctuations operate on that manifold and reduce the degeneracy.
Before getting into a quantitative analysis, it is worth pointing
out qualitative arguments towards an order out of disorder
mechanism.
The main ingredient for fluctuations to be efficient resides in
the spectrum of the system~: if two degenerate phases have
different spectra and in particular, if one has softer modes, it
is common that this one is selected at low temperatures, where
entropic effects and/or quantum fluctuations may take place.
The measure of the ground state manifold also plays an important
role and should be sufficiently small.
Schematically, if one phase is to be selected by fluctuations but
is "drowned" in a huge ground state manifold, the system cannot
statistically pick up that phase and fluctuations will therefore
not operate.
In the present case, as it is shown hereafter, both preceding
conditions are fulfilled : there are obviously peculiar points
expected to have spectra with softer modes (zero modes actually)
and the ground state manifold, because it is one dimensional, has
a null measure in configuration space and allows order out of
disorder to be efficient, both at thermal and quantum levels.

\subsection{Thermal fluctuations}
\label{subsec:thermal}

\begin{figure}[]
\vskip0.25cm
\begin{center}
\includegraphics[width=5.2cm,angle=-90]{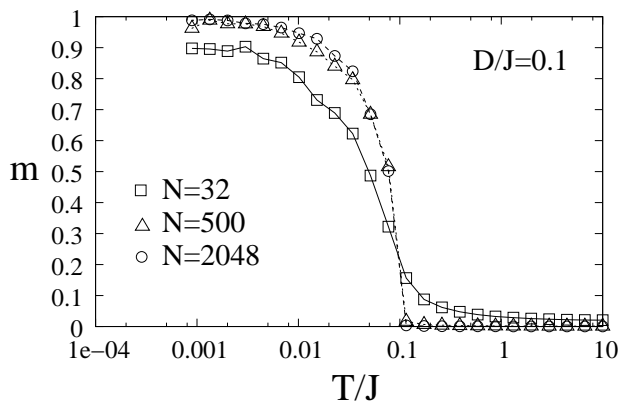}
\end{center}
\vskip0.25cm \caption{Thermal average of $m$ (see
Eq.~\ref{eq:define-M}) as a function of temperature for three
lattice sizes, $N=32, 500$ and $2048$, and for $D/J=0.1$.
} \label{fig:MdeT-D/J=0.1}
\end{figure}

\begin{figure}[]
\vskip0.25cm
\begin{center}
\includegraphics[width=5.2cm,angle=-90]{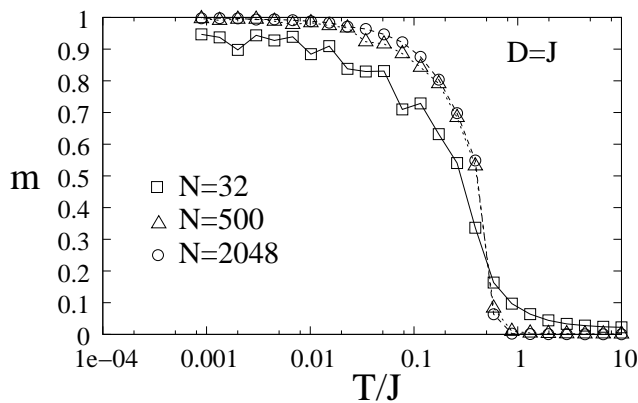}
\end{center}
\vskip0.25cm \caption{Thermal average of M (see
Eq.~\ref{eq:define-M}) as a function of temperature for three
lattice sizes, $N=32, 500$ and $2048$, and for $D/J=1.0$.
} \label{fig:MdeT-D/J=1.0}
\end{figure}

As emphasized in the previous paragraph we can guess that the
peculiar points of the configuration space around which thermal
fluctuations could be efficient are the above mentioned 6 points
because they are the only one within the ground state manifold
having two obvious "escape lines" e.g, two zero energy modes.
It is therefore natural to define a parameter which is maximum and
equal to one only when spin configurations are among the previous
six.
In each of these configurations, spins all belong to the same
plane, either $xy$, $yz$ or $zx$.
When they belong to the $xy$ plane, spins 1 and 4 are collinear as
are spins 2 and 3, with the 1-4 pair being perpendicular to the
2-3 pair (see figure \ref{fig:fig7}).
The same occurs for the two other planes.
Therefore, for each plane selection, a global index $\alpha$ can
be defined to label a vector $e_i^{\alpha}$ on each site of the
lattice, so that the quantity
\begin{equation}
m_{\alpha} = \frac{1}{N} \sum_i^N \vec{S}_i . \vec{e}_{\alpha}
\end{equation}
\noindent equals $\pm 1$ for the two time reversal related phases of
each plane selection $\alpha = xy, yz, zx$.
If the spins belong to the $xy$ plane, they point along the
$(1,-1,0)$ direction, thus $e_{xy} = \frac{1}{\sqrt{2}}(1,-1,0)$
and similarly for the two other planes.
Defining then
\begin{equation}
\label{eq:define-M} m = \max_{\alpha} \left( m_{\alpha}^2 \right)
\end{equation}
\noindent allows for characterizing any spin configuration and its
proximity to one of the six peculiar phases because $M=1$ for
these configurations and these configurations only.
In order to test wether thermal fluctuations may entropically
select peculiar phases of the ground state manifold, we have
performed classical monte carlo simulations of finite size
lattices with periodic boundary conditions.
The clusters we investigated had 32, 500 and 2048 sites.
For each simulations, we used a single flip metropolis algorithm
with a modified update so that the acceptance rate stayed roughly
above 40\%, associated to a local rotation of each spin around its
local molecular field.
During the simulation, the autocorrelation time is calculated on the
fly in order to adapt the number of monte carlo steps between two
measures and ensure that measures are uncorrelated.
Results are reported in Fig~\ref{fig:MdeT-D/J=0.1} and
\ref{fig:MdeT-D/J=1.0} for respectively $D/J = 0.1$ and $D/J = 1$.
The measure of $\left< m \right>$ as a function of the temperature
clearly indicates that this parameter is an order parameter,
although we cannot discuss the order of the transition so far.

Both cases support that at low temperature, the system orders in
one of the 6 previously described states as was expected from
qualitative arguments.
Moreover, it can be seen that $T_C$ is of the order of $D$,
which confirms that DMI are responsible for this ordering.

\subsection{Quantum fluctuations : semiclassical approach}
\label{subsec:quantum}

\begin{figure}[]
\begin{center}
\includegraphics[width=8.2cm,angle=0]{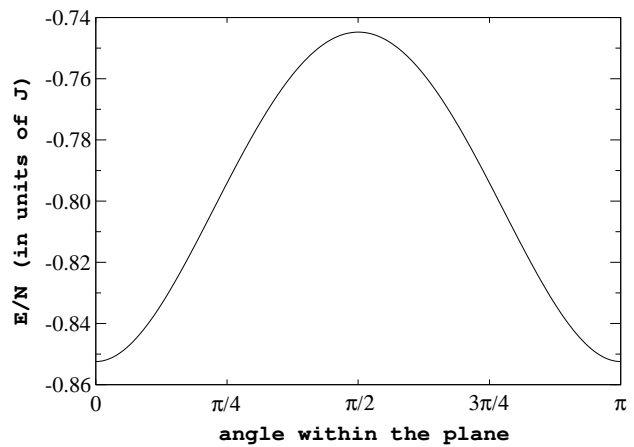}
\end{center}
\caption{Renormalized energy per site as a function of the angle
$\alpha$, for planar phases [all planes are equivalent], with
$D/J = 1$.
$\alpha = 0$ and $\alpha = \pi$ correspond to one time reversal
related pair of states belonging to the
$\mathbb{Z}_3\times\mathbb{Z}_2$ phases.
} \label{fig:EdePhi-D/J=1.0}
\end{figure}

\begin{figure}[]
\begin{center}
\includegraphics[width=8.2cm,angle=0]{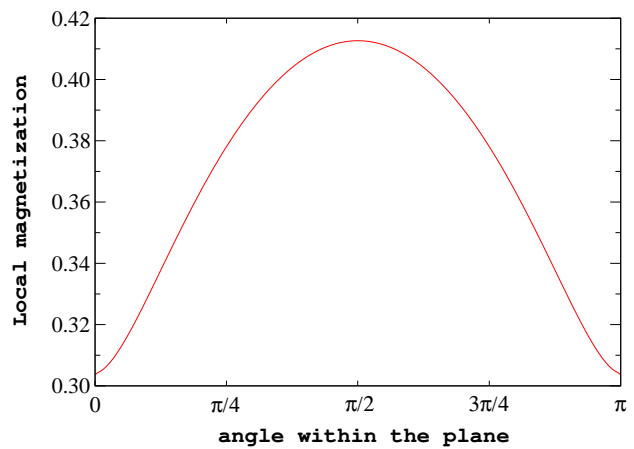}
\end{center}
\caption{Renormalized magnetization for $S=1/2$ as a function of
the angle $\alpha$, for planar phases, with $D/J = 1$.
$\alpha = 0$ and $\alpha = \pi$ correspond to one time reversal
related pair of states belonging to the
$\mathbb{Z}_3\times\mathbb{Z}_2$ phases.
At these two points, fluctuations have softer spectra, hence local
magnetization is much more renormalized.
} \label{fig:MdePhi-D/J=1.0}
\end{figure}

The same investigation has been done for quantum fluctuations.
Contrary to the classical case, it is not necessary to guess which
points will play a particular role.
This is because it is possible to compute at zero temperature the
energy renormalized by quantum fluctuations for all phases of the
ground state manifold.
To do so, we have performed a linear spin wave expansion of the
Hamiltonian around each state of the ground state manifold, e.g,
around each state parametrized by the angle $\varphi$.
The resulting bosonic hamiltonian is then diagonalized which
allows to obtain the normal modes and calculate the renormalized
energy and magnetization by quantum fluctuations.
The first result is that the $\mathbb{Z}_3\times\mathbb{Z}_2$
phases are unambiguously selected by quantum fluctuations, as
shown in Fig.~\ref{fig:EdePhi-D/J=1.0}.
For planar phases, they correspond to $\alpha = 0$ or
$\alpha = \pi$, $\alpha$ being the angle of the spins with one
of the 6 peculiar phases (2 for each plane).
The same calculation was performed for conic phases (indexed by
$\varphi$), confirming that the 6 phases are selected by quantum
fluctuations.
This selection is also illustrated by the computation of the
renormalized magnetization, as shown in
Fig.~\ref{fig:MdePhi-D/J=1.0}.
As confirmed by the calculation of the energy, there are softer
modes near the 6 selected phases.
Consequently, transverse fluctuations are much stronger giving a
larger renormalization of magnetization around these phases.
\begin{figure}[]
\begin{center}
\includegraphics[width=8.2cm,angle=0]{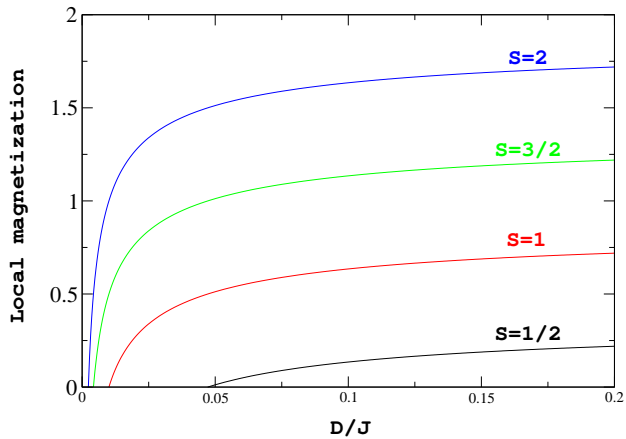}
\end{center}
\caption{Renormalized magnetization of one of the symmetry related
$\mathbb{Z}_3\times\mathbb{Z}_2$ phases, as a function of $D/J$
for $S=1/2, 1, 3/2$ and $2$.
For $S \ge 1$, a DM interaction of 1 percent is sufficient to
stabilize a Néel like long range order.
From that point of view, DM interactions are very efficient in
destroying the classical spin liquid behavior and driving this
model to a classical behavior.
} \label{fig:MdeD/J}
\end{figure}
Wether the magnetization, which is zero for $D/J=0$, could be
stabilized by taking DM's interactions into account is also of
interest because it quantifies how easily this model becomes
"classical".
We have therefore computed for one of the six selected phases, the
value of the quantum renormalized magnetization as a function of
$D/J$.
Results are reported in Fig.~\ref{fig:MdeD/J}.
It is worth noting that apart from the case $S=1/2$, a very small
value of $D/J$ ($D/J \gtrsim 0.01$) stabilizes the magnetic ground
state of the pyrochlore antiferromagnet.
Furthermore, we can quantify how efficient DM's interactions are
by fitting the gain in energy as a function of $D/J$.
For small values of $D/J$, quantum fluctuations contribute to the
renormalization of the energy as $\sqrt{D/J}$.
This is an important point as discussed in the next section.

\section{ROLE OF ANISOTROPIC SYMMETRIC LIKE EXCHANGE}
\label{sec:anisotropic}

So far, only DM's interactions have been taken into account.
Whatever the type of fluctuations, thermal or quantum, the order
out of disorder mechanism is efficient and reduces the continuous
degeneracy to a discrete one.
But is this selection robust vis à vis next order (in spin orbit
coupling) interactions present in the Hamiltonian, the anisotropic
symmetric exchange?
Unfortunately, we must lose generality to address this question
because many kind of symmetry allowed interactions are possible.
It is nevertheless possible to argue quantitatively on that
matter.
In order to be specific, we choose to consider interactions that
keep all symmetries of the pure pyrochlore lattice.
They correspond to nearest neighbour truncated dipolar
interactions\cite{footnote1},
\begin{equation} \label{eq:an-exchange}
K  \,  \frac{({\hat n}^{\alpha}\cdot R_{ij}^{mn}) ({\hat
n}^{\beta}\cdot R_{ij}^{mn})}{|R_{ij}^{mn}|^5} \; .
\end{equation}
where indices $i, j$ are related to the fcc Bravais lattice while
indices $m, n$ correspond to one of the four sites of the
tetrahedral unit cell.
The set of unit vectors ${\hat n}^{\alpha}$ is {(1,0,0), (0,1,0),
(0,0,1)}.
We allow the sign of $K$ to vary as there is no obvious reason why
one peculiar type, $K>0$ or $K<0$, should prevail.
The full hamiltonian we are left with reads now,
\begin{equation}
\mathcal{H} = - \frac{1}{2}\sum_{i,j} \sum_{m,n} \sum_{\alpha,
\beta}  {\mathcal{J} \left( R_{ij}^{mn} \right)}^{\alpha \beta}
 {S}_{i}^{m,\alpha}
{S}_{j}^{n,\beta}
\end{equation}
where ${\mathcal{J} \left( R_{ij}^{mn} \right)}^{\alpha \beta}$
are the coefficients of the coupling matrix $\mathcal{J} \left(
{\bm q} \right)$ which contains the isotropic symmetric exchange
$- J S_i . S_j$, the antisymmetric exchange $D_{ij} . S_i \times
S_j$ and the anisotropic symmetric exchange $K$ (see
Eq.~\ref{eq:an-exchange}).
Stabilized low temperature phases are investigated through a mean
field analysis.
The hamiltonian is rewritten in reciprocal space using the
following transformations,
\begin{eqnarray}
\label{eq:eq-FTm} S_{i}^{m,\alpha} = \frac{1}{\sqrt{N}} \sum_{\bm
q} S_{\bm q}^{m,\alpha} \, e^{-\imath {\bm q} \cdot {\bm
R}_{i}^{m}},
\\ {\mathcal{J} \left( R_{ij}^{mn} \right)}^{\alpha \beta} =
\frac{1}{N} \sum_{\bm q} {\mathcal J}_{mn}^{\alpha \beta}({\bm q})
\, e^{\imath {\bm q} \cdot {\bm R}_{ij}^{mn}}, \label{eq:eq-FTJ}
\end{eqnarray}
where $N$ is the number of Bravais lattice points.
The resulting interaction matrix ${\mathcal J}({\bm q})$ is a $12
\times 12$ non diagonal hermitian matrix.
Hence, to completely diagonalize ${\mathcal J}({\bm q})$ one must
transform the ${\bm q}$-dependent variables, ${\bm S}_{\bm q}^m$,
to normal mode variables. In component form, the normal mode
transformation is given by
\begin{equation}
\label{eq:eq-nmodes} S_{\bm q}^{n,\alpha} =
\sum_{p=1}^{4}\sum_{\gamma=1}^{3} U_{n,p}^{\alpha,\gamma}({\bm q})
\phi_{\bm q}^{p,\gamma},
\end{equation}
where the indices ($p,\gamma$) label the normal modes ($12$ for
Heisenberg spins), and $\{\phi_{\bm q}^{p, \gamma}\}$ are the
amplitudes of these normal modes. $U({\bm q})$ is the unitary
matrix that diagonalizes ${\mathcal J}({\bm q})$ with eigenvalues
$\lambda({\bm q})$.
Finally, the mean-field free energy to quadratic order in the
normal modes reads, up to an irrelevant
constant\cite{Enjalran2004},
\begin{equation}
{\mathcal F}(T) = \frac{1}{2} \sum_{{\bm q}, p, \gamma} (nT -
\lambda_{p}^{\gamma}({\bm q})) |\phi_{\bm q}^{p, \gamma}|^2 ,
\label{eq:eq-fmf}
\end{equation}
where ${\mathcal F}(T)$ is the mean-field free energy per unit
cell, $T$ is the temperature in units of $k_{\rm B}$, and $n=3$
for Heisenberg spins.
Therefore, the mean-field low temperature phase is defined by the
corresponding wave vector ${\bm q}_{\rm ord}$ associated with the
extremal eigenvalue ${\max}_{p, \gamma, {\bm q}} \left(
\lambda_{p}^{\gamma}({\bm q}) \right)$.
When $K=0$, it was shown in preceding sections that ${\bm q}_{\rm
ord} = 0$ and that the ground state manifold is continuously
degenerate.
Including non zero $K$, whatever its sign, the degeneracy is
lifted already at the mean field level.
The new ground states are, for both positive and negative $K$,
${\bm q} = 0$ slightly distorted version of the $\mathbb{Z}_3 \times
\mathbb{Z}_2$ states.
The energy decrease, for small $K$, is always quadratic, $\Delta E
\propto K^2$.

At this point, we can discuss wether anisotropic symmetric
interactions may interfere with quantum fluctuations induced by
DM's interactions.
The first remark is that at the mean field level, $K$ breaks the
degeneracy but select phases which are continuous deformation of
quantum selected phase.
Therefore, we expect that taking into account anisotropic exchange
would not change the behavior of this system.
The second point is related to how strong $K$ is efficient in
selecting a particular phase.
Suppose that the mean field selected phase is {\it not} one of the
$\mathbb{Z}_3 \times \mathbb{Z}_2$ phases or a slightly distorted
version of those.
Because the energy gain is quadratic in $K$, it appears that
quantum fluctuations are much more efficient, as they induce an
energy gain proportional to $\sqrt{D}$ (see section
\ref{subsec:quantum}).
This clearly shows that even if anisotropic interactions are taken
into account, we expect a lifting of degeneracy corresponding to
the $\mathbb{Z}_3 \times \mathbb{Z}_2$ phases or slightly
distorted versions of those.
The only mechanism which could work against such scenario would be
that quantum fluctuations around $K$ selected phases be larger
than $\sqrt{D}$.
Because $K \propto D^2$, this would mean that one should expect
the energy gain induced by $K$-like interactions to behave like
$\sqrt[4]{K}$, which is highly unprobable.

\section{CONCLUSION}
\label{sec:conclusion}

This works investigates the role of thermal and quantal
fluctuations on the antisymmetric pyrochlore antiferromagnet.
When the symmetry type of the allowed DMI leads to a one
dimensional degenerate ground state, it is shown that these two
types of fluctuations are efficient in reducing the degeneracy and
drives the model to the same ordered ground state with a discrete
global degeneracy of the $\mathbb{Z}_3 \times \mathbb{Z}_2$ type.
Including higher order terms in the hamiltonian like the
anisotropic symmetric exchange is likely to leave thermal and
quantal fluctuations dominate the low temperature behavior of this
system.
It is therefore concluded that whatever the type of symmetry
allowed DMI, direct or indirect, they should drive the pyrochlore
Heisenberg antiferromagnet to a low temperature ordered phase.
When the model is essentially classical, the critical temperature
is of order {\bf D}.
If quantum fluctuations are present, the value of the ordering
temperature  results from a balance between the strength of the
DMI and the magnitude of the fluctuations.
As we have shown that for moderately large DMI the local
magnetization is quite stable, the value of the ordering
temperature should not be very different from the one of the
classical case.
Finally, it is worth pointing out that most of studied pyrochlore
compounds are Rare Earth oxides.
It is therefore the next natural step to investigate the mechanism
of DMI within 5f alloys and determine whether those interactions
could be in some compounds large enough to drive the low
temperature behavior of these frustrated systems.
%

%\bibliography{}

\begin{thebibliography}
\bibitem{footnote1}The symmetric part of the dipolar-like
interaction can be absorbed in the usual symmetric exchange.
\end{thebibliography}
\end{document}